# A Miniaturized Prototype of Resonant Banana-Shaped Photoacoustic Cell for Gas Sensing

A.L.Ulasevich, A.V.Gorelik, A.A.Kouzmouk, V.S.Starovoitov*
B.I.Stepanov Institute of Physics, NASB, Nezavisimosti ave. 68, 220072 Minsk, Belarus
* Corresponding author. E-mail: vladstar@dragon.bas-net.by (V. Starovoitov)

**ABSTRACT**

A resonant photoacoustic cell intended for laser-spectroscopy gas sensing is represented. This cell is a miniature imitation of a macro-scale banana-shaped cell developed previously. The parameters, which specify the cavity shape, are chosen so as not only to provide optimal cell operation at a selected acoustic resonance but also to reduce substantially the cell sizes. A miniaturized prototype cell (the volume of acoustic cavity of ~ 5 $mm^3$) adapted to the narrow diffraction-limited beam of near-infrared laser is produced and examined experimentally. The noise-associated measurement error and laser-initiated signals are studied as functions of modulation frequency. The background signal and the useful response to light absorption by the gas are analyzed in measurements of absorption for ammonia in nitrogen flow with the help of a pigtailed DFB laser diode oscillated near a wavelength of 1.53 µm. The performance of prototype operation at the second longitudinal acoustic resonance (the resonance frequency of ~ 32.9 kHz, $Q$-factor of ~ 16.3) is estimated. The noise-limited minimal detectable absorption normalized to laser-beam power and detection bandwidth is ~ $8.07 \cdot 10^{-8}$ $cm^{-1}$ W $Hz^{-1/2}$. The amplitude of the background signal is equivalent to an absorption coefficient of ~ *$2.51 \cdot 10^{-5}$* $cm^{-1}$. Advantages and drawbacks of the cell prototype are discussed. Despite low absorption-sensing performance, the produced miniaturized cell prototype shows a good capability of gas-leak detection.

## 1. Introduction

Laser photoacoustic spectroscopy is an efficient technique to be applied to local non-contact analysis of trace amounts of various chemical compounds in gas media [1-4]. The principle of the technique is based on measuring the amplitude and phase for an acoustic pressure oscillation (a so-called photoacoustic response or signal) arising due to absorption of a modulated laser beam by molecules of gas inside a photoacoustic cell. The response amplitude is proportional to the beam power and absorption coefficient. The photoacoustic gas detection is realized with an enhanced sensitivity if the modulation frequency coincides with an acoustic resonance of the internal cell cavity. The resonant photoacoustic technique based on infrared lasers is distinguished by a high sensitivity (the minimal detectable absorption of $10^{-10}$ $cm^{-1}$ can be attained at a time resolution of a few seconds) and capability to recognize reliably a large number of chemical compounds [5-8]. The technique has found successful exploitation for a large number of practical applications [9-13].

A promising line of development for the photoacoustic technique is associated with creating a miniature resonant cell [14,15]. According to theoretical estimations (see, for instance, [14, 16, 17]), the amplitude of photoacoustic response can be increased with reducing the cell sizes. The miniature high-sensitivity photoacoustic cell is a valuable facility in order to analyze chemical compounds to be emitted by individual small-sized objects at an extremely low emission rate. The compact photoacoustic gas sensor can find an successful exploitation as a gas-leak detector. A crude estimation shows that the gas-leak-detection sensitivity of a laser photoacoustic sensor equipped with a resonant cell, the cavity volume of which is ~ 1 $mm^3$, can be some hundred times higher in comparison to the commercial mass-spectrometer-based leak-detection systems [18]. The photoacoustic leak detector can be applied to *in situ* localization of leak for a large number of gases to be emitted in atmospheric air. The most appropriate light sources for the compact photoacoustic gas sensor are miniature single-mode semiconductor lasers operated in the near- or mid-infrared wavelength region. They are laser diodes (the oscillation wavelength region of 0.4 - 4 µm) or quantum cascade lasers (4 - 20 µm) [19-21].

In practice, there are a few efficient approaches to miniaturize the resonant photoacoustic cell. The quartz-enhanced photoacoustic spectroscopy (QEPAS) is a well-developed way to the miniature gas sensor [22]. Instead of a gas-filled resonant acoustic cavity, the sound energy is accumulated in a high-Q quartz crystal frequency standard. Usually, the standard is a quartz tuning fork with an acoustic resonant frequency of 32 kHz in air. A theoretical model (that enables the detected piezoelectric signal to be expressed in terms of optical, mechanical, and electrical parameters of the QEPAS sensor) has been developed [23]. A great success is achieved in QEPAS-based gas detection with the help of different infrared laser systems (including laser diode, quantum cascade lasers and optical parametric oscillators) [24-28]. The volume of acoustic cavity for the developed cells reaches down to a few cubic centimeters. The QEPAS-based experiments demonstrate that the high-sensitivity photoacoustic detection can be realized at ultrasonic modulation frequencies if the rates of intra- and intermolecular collisional vibration-vibration (VV) or vibration-translation (VT) energy redistribution for the detected species are high compared to the modulation frequency.

Another approach to the cell miniaturization implies a traditional design of photoacoustic cell: through a small hole in the cell shell, an acoustic sensor (for instance, a condenser microphone or an ultrasonic transducer) registers response to the light beam modulated with the frequency of an acoustic resonance of cell cavity. The cell design must be thoroughly optimized in order to provide the best gas-detection performance for the selected acoustic resonance. The most accurate way of optimization can be performed with the help of a numeric simulation, which gives the three-dimension amplitude distribution for the resonant acoustic standing waves and estimates correctly the spatial acoustic signals to be generated inside the cell. Our recent study demonstrates that such a numeric-simulation-based optimization allow one to minimize the parasite acoustic signals (the noise to be initiated by external acoustic disturbances, the background due to absorption and reflection of laser beam by the cell windows), which can play a great negative role for the small-sized cells [29,30]. The standard finite-element method, which gives the direct numerical solution of acoustic Helmholtz equation, is a reliable well-proven technique to be applied for optimizing the design of photoacoustic cell [31-34]. The photoacoustic cell can be also miniaturized down to microelectromechanical-system dimensions by a trivial size-scaling procedure applied to a well-developed macro-scale cell [35-37].

Here we apply the size scaling in order to produce a miniaturized analogue of a so-called banana-shaped cell developed previously [38-40]. The banana-shaped cell (named after its cavity shape) is designed for operation at a longitudinal acoustic resonance. The optimum design of the cell is found by numerical simulation (using an electric equivalent model) of acoustic properties for various cell geometries and examined in experiments [38]. The cell is specified by a high gas-detection sensitivity and moderate background signals. The measured microphone-noise-limited minimum detectable absorption coefficient is approximately $2 \cdot 10^{-10}$ cm$^{-1}$ W Hz$^{1/2}$. The background signal (its amplitude corresponds to an absorption coefficient $\alpha_{bg,macro} \approx 2 \cdot 10^{-8}$ cm$^{-1}$) is low compared to the microphone noise if the applied laser power is below 10 mW. The volume $V_{macro}$ of acoustic cavity for the cell is 108 cm$^3$.

In a sense, our miniature photoacoustic cell-analogue is an imitation of the macro-scale cell [38]. Like the cell [38], our cell is designed for operation with a collimated linear-polarized light beam. The parameters, which specify the cavity shape of the cell, are chosen so as not only to provide the optimal cell operation at a selected resonance but also to reduce essentially the cell sizes. In order to realize the cavity sizes as small as possible the cell design must be fitted to a substantially narrow low-divergence light beam. The goal of the work is to clarify advantages and drawbacks of such a trivial imitation-based approach to the miniaturization of resonant photoacoustic cell. In the paper a cavity design for the photoacoustic banana-shaped cell is represented. An experimental examination of cell performance is made for a miniature cell prototype, which is adapted to the narrow diffraction-limited light beam to be generated by a near-infrared semiconductor laser. In the examination we estimate the frequency spectrum of measurement error associated with acoustic-sensor noise and test the acoustic isolation of the prototype from the environment. Then we analyze the amplitude-frequency dependence for photoacoustic signals (they are the window background and useful response to light absorption by the gas) to be initiated by the laser beam inside the prototype cell. The sensitivity of photoacoustic gas detection for the prototype is estimated in terms of the minimal detectable absorption coefficient. Advantages and drawbacks of the produced cell prototype are discussed.

## 2. Miniaturized cell prototype

### 2.1. Basic design of cell cavity

Here we describe a design of acoustic cavity for the resonant photoacoustic cell. The cell is intended for detection of gases with the help of a collimated and linearly polarized light beam. As for the cell [38], the acoustic cavity of our cell consists of three adjacent cylindrical parts (one central and two lateral cylinders) and resembles a banana shape. The cross-section diameters for the central and lateral cylinders are accepted to be identical and equal to $D$. The symmetry axes of all three cylinder cavities are located on the plane $OO'P$ formed by the optical axis $OO'$ and the electric polarization vector $P$ for the light beam. Figure 1 shows a section of the acoustic cavity of cell by the plane $OO'P$.

The light beam is accepted to go through the central cylinder part along the optical axis $OO'$, which coincides with the axis of cylindrical symmetry for this cavity part. The points $O$ and $O'$ are points of intersection of the axis $OO'$ and the internal surfaces of optical front- and back-end windows of the cell. In order to minimize the parasite reflection of light beam on the window surfaces the front- and back-end windows are mounted at the Brewster angle $\Theta_B$ relative to the axis $OO'$ ($\Theta_B$ is an angle between $OO'$ and the normal to the window surface). The distance $L_C$ between the points $O$ and $O'$ is accepted as a length of the central cylinder part.

The lateral cylinder parts are identical in the sizes. The parts are located symmetrically relative to the central part. The symmetry axes of these parts go through the point $O$ or $O'$. Each of these axis is inclined at the angle $\pi - 2\Theta_B$ relative to the axis $OO'$. Owing to the incline, the parasite beam reflection on the surfaces of the back-end window is withdrawn from the cell cavity through one of the additional optical windows mounted in the ends of the lateral cylinders. For each lateral cylinder parts, the cylinder length $L_{Lat}$ is a distance between the point $O$ (or $O'$) and internal surface of the end window along the cylinder symmetry axis.

Like the cell [38], our photoacoustic cell is optimized for operation at the second longitudinal acoustic mode (denoted as a $v_2$ mode) of the cell cavity. For minimizing the window background (a photoacoustic response arisen in the cell due to absorption of the light beam in the cell windows), we accept (as in [38]) that $L_{Lat} = L_C/2$. At this length proportion the nodes surfaces of $v_2$ mode are planes, which coincide with the intersections of the central and lateral cavity parts. The nodes of $v_2$ mode are located in the points $O$ and $O'$ (that is, in the vicinity of the cell windows). Inlet and outlet gas holes are located near the node surfaces of $v_2$ mode. Such a location of the gas holes is intended to reduce the negative influence of holes on the acoustic $Q$-factor and, simultaneously, to isolate the measurement from external acoustic noise for the $v_2$ mode. Ducts, which connect the inlet/outlet gas holes to nipples, are directed up perpendicularly to the plane $OO'P$. The nipples are adapted to flexible gas tubing. The photoacoustic response is registered by an acoustic sensor $T$ located near the midpoint of the central cavity part.

*2.2. Cell prototype*

The presented design of cell cavity is implemented in a produced miniaturized prototype of photoacoustic cell. The parameters, which specify the cavity shape for the prototype, are chosen so as not only to provide an optimal cell operation at the acoustic $v_2$ mode but also to reduce essentially the cell sizes. The prototype is adapted to the narrow diffraction-limited light beam to be generated by a near-infrared semiconductor laser. The prototype shell is made of brass. The cross-section diameter $D$ for the central and lateral cylinders is 0.8 mm. The lengths ($L_C$ and $L_{Lat}$) of the central and lateral cylinder parts are, correspondingly, 5 and 2.5 mm. The optical windows are made of $CaF_2$. The angle $\Theta_B$ is accepted to be equal 55 degree. This angle can be accepted as the Brewster angle for $CaF_2$ over a broad near-infrared wavelength range. An ultrasonic MEMS-based transducer Knowles Acoustics SPM0204UD5 (sensitivity ~ -47 dB, signal-to-noise ratio ~ 59 dB) is applied as the acoustic sensor T mounted in the cell shell. The transducer is connected to the acoustic cell cavity by a duct (the duct diameter of 0.7 mm, the duct length of 1 mm). The cross-section diameter for the inlet and outlet holes in the cell shell is 0.3 mm. The volume $V$ of acoustic cavity for the prototype is approximately 4.6 mm$^3$. The total cell weight (including the cell shell, windows, nipples, transducer and electric wiring) is 9.8 g. According to our estimation, the eigen-frequency of $v_2$ mode for the cavity prototype is approximately 33.97 kHz. The photo of the prototype cell is represented in Figure 2.

## 3. Noise-associated measurement error

### 3.1. Procedure of measurement-error estimation

In the experiment, we estimate a measurement error for the detected photoacoustic response due to noise to be generated by the transducer inside the prototype cell in the absence of light beam. We apply an error-estimation technique, which is similar to a procedure described in [30].

The prototype cell is inserted into a capsule, which provides a reliable sound isolation from the noise produced in our laboratory room (see Figure 3). The capsule design allows connection of the cell cavity with the environment through only inlet and outlet flexible gas ducts of the inner diameter of ~ 1 mm. The cell cavity is filled with conditioned laboratory air. The open inlet and outlet gas ducts of capsule allow the air flow through the cell at a flow rate (~ 0.5 cm$^3$/min), which provides a perfect gas renewal inside the cell for the time ~ 0.6 sec.

The voltage signal $S_t$ from the cell transducer is amplified by a frequency-selective low-noise amplifier (selective nanovoltmeter type 273, Unipan) and stored by a HS3 digital oscilloscope connected to a personal computer. The transducer signal is stored by the oscilloscope as a time-sample signal realization over a fixed time interval $\tau_I \approx 0.262$ s. The photoacoustic response is determined as a Fourier transform $S_f$ of the transducer signal $S_t$. The quantity $S_f$ is a complex-valued function of frequency $f$ calculated with the help of a fast-Fourier-transform procedure performed for an individual time-sample signal realization stored in the oscilloscope. The time of signal averaging $\tau_{avr}$ for the quantity $S_f$ is equal to the time of signal realization $\tau_I$. The value of error for each frequency component of quantity $S_f$ is determined as a bandwidth-normalized standard root-mean-square deviation of $S_f$:

$$\sigma_f = \tau_{avr}^{1/2} (<S_f S_f^*> - <S_f><S_f^*>)^{1/2}. \qquad (1)$$

The symbol $<\ldots>$ means the averaging over an ensemble of the time-sample signal realizations. The number of the signal-sample realizations used for the ensemble-averaging procedure is not less than 1000. We evaluate the noise-associated measurement error for each frequency component of photoacoustic response and analyze the frequency spectrum of $\sigma_f$.

### 3.2. Frequency spectrum of noise-associated measurement error

The frequency spectrum of standard deviation is analyzed at different levels of transducer noise. We assume that the cell cavity can be coupled acoustically with the environment through the inlet and outlet gas ducts. Therefore, in the estimation we take into account a possible manifestation of acoustic disturbances produced outside the cell by a source.–The measurement error of transducer signal is estimated at three distinct strengths of external-noise effect on the cell transducer. The obtained frequency spectra of the deviation $\sigma_f$ are represented in Figure 4 for the frequency range from 0.1 to 77 kHz.

A negligibly small influence of external noise on the cell transducer is realized when the inlet and outlet gas ducts are closed and the cell cavity is acoustically isolated from the environment. A regime of strict silence is kept in the laboratory room. The measurement error $\sigma_f^{(0)}$ to be obtained at such an acoustic-isolation regime corresponds to the minimal error level attainable by the transducer signal. The observed deviation $\sigma_f^{(0)}$ is shown in Figure 4 to be a slowly varying function of $f$. Over the frequency range from 3 to 70 kHz, the magnitude of $\sigma_f^{(0)}$ is in the interval of values from 0.15 to 0.2 µV/Hz$^{1/2}$. We associate this measurement-error value with inherent electric noise of the applied transducer.

A moderate action of external noise is realized with the opened inlet and outlet gas ducts. A standard day-to-day noise level is kept in the laboratory room. In the room there are no specially-designed sources of external disturbances. Figure 4 shows a frequency spectrum of the deviation $\sigma_f^{(1)}$ obtained at this regime. The quantity $\sigma_f^{(1)}$ is shown in the figure to be significantly high in comparison with $\sigma_f^{(0)}$ at low frequencies ($f < 5$ kHz). At higher frequencies ($f > 5$ kHz) the observed frequency-dependence of $\sigma_f^{(1)}$ is identical to that of the quantity $\sigma_f^{(0)}$. At a frequency $f = 32.9$ kHz, the quantities $\sigma_f^{(0)}$ and $\sigma_f^{(1)}$ are equal to a noise level $\sigma_{f2} \approx 0.082$ µV/Hz$^{1/2}$.

A strong effect of external noise on the measurements is implemented at operation of a specially-designed source of external acoustic disturbances. In our measurements, the disturbances are generated by a compressed air flow blown through a nozzle. Figure 4 demonstrates a typical frequency spectrum of the nozzle-disturbance-initiated measurement error obtained for a free-spaced ultrasonic sensor SPM0204UD5 (a transducer identical to the sensor mounted inside the cell) located immediately in front of the nozzle. According to this spectral data, acoustic disturbances to be produced by the nozzle may lead to a significant noise increase for the cell-unmounted transducer over a broad frequency scale ranging from 5 to 70 kHz.

We evaluate the strong effect of the nozzle-produced acoustic disturbances on the transducer mounted inside the prototype cell at the opened inlet/outlet gas ducts. The nozzle is placed near the suction hole of inlet duct of the capsule as it represented in Figure 3. Figure 4 shows a frequency spectrum of the deviation $\sigma_f^{(s)}$ observed at this regime. The obtained quantity $\sigma_f^{(s)}$ is significantly high compared to $\sigma_f^{(0)}$ for a low- and high-frequency ranges (correspondingly, at $f < 34$ kHz and $f > 51$ kHz). Within the frequency range from 34 to 51 kHz, the observed deviation $\sigma_f^{(s)}$ is close to the quantities $\sigma_f^{(0)}$ and $\sigma_f^{(1)}$.

Hence, our experiment shows that, for a restricted frequency range, the measurement error is not affected by the external disturbances. Regardless of the strengths of applied disturbances, the standard deviation does not exceed the value of ~ 0.17 µV at frequencies $f$ from 35 to 50 kHz. As in [30], we associate this prototype-cell immunity to the disturbances with arrangement of the inlet and outlet gas holes drilled in the cell shell. Notice that such an acoustic isolation for a restricted frequency range is efficient if the external disturbances are applied to the cell through only the inlet and outlet gas ducts.

## 4. Laser-initiated signals

### 4.1. Experimental setup

The laser-initiated photoacoustic signals arisen inside the presented photoacoustic prototype cell are analyzed by experimental way. The used experimental setup is shown in Figure 5. In the experiment, the prototype is examined with the help of a collimated light beam of near-infrared laser diode. The light beam is generated by a current-modulated laser source. The source is a fiber-pigtailed single-mode distributed-feedback laser diode (D2547PG57, Agere Laser 2000) operated near a wavelength of 1.53 µm. The current for the laser diode is supplied by a Thorlabs current controller LDC 202C. The current modulation is performed with the help of a TTL-like reference signal directed from a Handyscope HS3 digital oscilloscope to the current controller. The frequency $f_m$ of reference-signal switching is accepted as a frequency of beam modulation. A Thorlabs TED 350 temperature controller is used in order to maintain the laser diode at a fixed temperature.

The laser beam is output from a PM-fiber and a Thorlabs CFC-2X-C collimator and directed through a half-wave quartz plate and the prototype cell to a power meter (Ophir PM 3A). The half-wave plate (it is optimized for operation near a wavelength of 1.535 µm) is used for fine orientation tuning of the beam polarization vector $P$ to the plane, which is formed by the symmetry axes of all three cylinder cavities of the prototype cell. The efficient cross-section diameter of the collimated laser beam, which passes the central cavity part of the prototype, is ~ 0.38 mm. The prototype cell is thoroughly adjusted in such a way as to provide the best transmission of the beam through the prototype. In the experiments, the laser-beam power $P_{on}$, which is measured by the power meter immediately after the optical back-end window of the prototype, is ~ 6.5 mW.

The voltage signal from the cell transducer is amplified by a frequency-selective low-noise amplifier (selective nanovoltmeter type 273, Unipan) and stored by a HS3 digital oscilloscope connected to a personal computer. The transducer signal is stored by the oscilloscope as a time-sample signal realization over a fixed time interval $\tau_1 \approx 0.262$ s. The reference signal from the oscilloscope is applied in order to synchronize the triggering of stored realizations. The photoacoustic response is determined as a Fourier transform $S(f_m)$ of the transducer signal $S_t$ for a frequency $f = f_m$ [41]. The quantity $S(f_m)$ is a complex-valued number calculated with the help of a fast-Fourier-transform procedure performed for an individual time-sample signal realization stored in the oscilloscope. In the study we analyze a quantity $<S(f_m)>_{(n)}$, which is an average of the measured response $S(f_m)$ over $n$ signal-sample realizations. The time of signal averaging $\tau_{avr}$ for such an averaged quantity is equal to $n\tau_1$.

Here, we analyze photoacoustic signals, which are arisen due to absorption of the laser beam by ammonia. The absorption spectrum for ammonia near a wavelength of 1.53 µm is a group of lines, which belong to overtone and combination band transitions [42]. The lines are overlapped at atmospheric pressures. All our measurements are performed at a fixed laser wavelength (1531.67 nm), which corresponds to a line-group peak to be observed in the absorption spectrum of ammonia. The laser diode is tuned accurately on this wavelength with the help of the temperature controller. According to [42,43], the ammonia absorption coefficient $\alpha^{(NH3)}$ for this laser wavelength at atmospheric pressure and room temperature (295 K) is 0.34 cm$^{-1}$atm$^{-1}$. We assume that the rates of VV/VT redistribution for the energy to be absorbed by ammonia are high compared to the modulation frequencies realized in the experiment.

All measurements in the experiment are made for flows of ultra-high purity nitrogen (99.9995 %) produced by a nitrogen generator (ANG250A, Peak Scientific Instruments LTD). Ammonia is admixed to the nitrogen flow at a concentration $C^{(NH3)} \approx 3954$ ppm. The ammonia is produced with the help of a calibrated permeation tube (IM 06-M-A2, Analitpribor) inserted in a leak-proof box. The rate of gas flow (0.52 cm$^3$/min) to be blown through the cell is maintained automatically with the help of a flow controller.

The wall adsorption can play a great negative role for detection of ammonia at a ppm-concentration level. Therefore, we make some arrangements which provide reliably the finishing of wall adsorption before the measurements. In order to accelerate the adsorption process, the box with ammonia tube and the cell prototype are connected by a short Teflon tube of small cross section (the length and internal diameter of the tube are, correspondingly, 30 cm and 1.2 mm). Before the measurement, all the gas line is kept at the required nitrogen-ammonia flow during a long time period sufficient for the absorption finish. Usually, this period is not shorter than 48 hours. All the measurements show good hour-by-hour and day-by-day reproducibility. The maximal amplitude of photoacoustic response associated with ammonia presence in the gas flow is reproduced within a relative error of 10 %.

The pressure and temperature of gas in the prototype cell is accepted to be close to typical parameters for the laboratory room (740 torr and 22 degrees Celsius). Notice that, in the experiments, no capsule is applied for acoustic isolation of the cell. Similar to our study of the moderate effect of external noise on the transducer signal, a standard day-to-day noise level is kept in the laboratory room. A more detailed description of the used experimental set-up is given in [44].

*4.2. Amplitude-frequency dependence of laser-initiated photoacoustic responses*

We analyze the photoacoustic response to laser-beam absorption inside the prototype cell as a function of modulation frequency $f_m$ for a frequency range from 10 to 60 kHz. The laser-initiated response is analyzed as a sum of components. The components are a parasite background signal (we associated this background, mainly, with absorption of laser beam in the cell windows) and a useful signal (a response to light absorption by gas inside the cell). In order to determine these signal components we perform the measurements at two distinct gas flows (a flow of pure nitrogen or a nitrogen flow admixed by ammonia) blown through the cell.

Obviously, the response to be detected at the pure-nitrogen flow is the background signal. In order to reduce the negative influence of the noise on the measurements and to provide a reliable determination of this signal, we perform the averaging of the signal over an ensemble of multiple realizations. We find the quantity $<S(f_m)^{(N2)}>_{(40)}$, which is the background signal averaged over *40* signal-sample realizations. According to our tentative estimation, the error of determination for this signal is ~ $2.5 \cdot 10^{-2}$ µV. The response, which is observed at the ammonia-containing nitrogen flow, is a sum manifestation of the background and useful signals. This response is found as a quantity $<S(f_m)^{(NH3)}>_{(4)}$ averaged over 4 signal-sample realizations. The measured amplitudes of the responses $<S(f_m)^{(N2)}>_{(40)}$ and $<S(f_m)^{(NH3)}>_{(4)}$ are represented in Figure 6 as functions of modulation frequency $f_m$.

According to the obtained data, the amplitude-frequency dependence of the background response $<S(f_m)^{(N2)}>_{(40)}$ exhibits no resonance peak for frequencies near the eigen-frequency of the second longitudinal acoustic mode. A clear minimum (~0.05 µV) for the amplitude $|<S(f_m)^{(N2)}>_{(40)}|$ is observed at a modulation frequency of ~ 40 kHz. In the presence of ammonia in the gas flow, the amplitude-frequency dependence of response is essentially transformed. Admixing ammonia to the flow leads to an increase in the response amplitude. The observed response amplitude $|<S(f_m)^{(NH3)}>_{(4)}|$ demonstrates clearly a resonance peak near the eigen-frequency of the $v_2$ mode. The peak amplitude is realized when the modulation frequency is close to a frequency $f_2 \approx 32.9$ kHz. Obviously, the observed peak is a manifestation of the resonance between the modulated laser beam and acoustic $v_2$ mode. For the resonance peak, the Q-factor $Q_2$ ($Q_2 = f_2/\Delta f_2$, the width $\Delta f_2$ is measured between the points where the amplitude is a $1/2^{1/2}$ value of the peak amplitude) is approximately 16.3. The resonance answers to the maximum portion of the useful signal in relation to the noise and background. Definitely, for an ammonia concentration $C^{(NH3)}$ (some thousands of ppm) to be applied in the measurements, the observed response $S(f_m)^{(NH3)}$ at $f_m = f_2$ can be accepted as a useful signal. The ultimate sensitivity of gas detection may be realized for the cell prototype when the modulation frequency $f_m$ is close to the resonance-peak frequency $f_2$.

## 5. Cell performance at acoustic resonance

In order to estimate the performance of the cell prototype operated in resonance with acoustic $\nu_2$ mode we evaluate background and useful photoacoustic signals to be generated at the modulation frequency $f_m = f_2$. In the evaluation, the experimental setup and measurement procedure are the same as in our study of laser-initiated signals (see the details in subsection 4.1). All measurements are done at a laser beam power of 6.66 mW.

Accurate measures of useful and background signals are the amplitudes and standard deviations for the responses $<S(f_2)^{(NH3)}>_{(1000)}$ and $<S(f_2)^{(N2)}>_{(1000)}$ averaged over a sufficiently long time scale $\tau_{avr} = 1000\tau_1 \approx 262s$. According to our measurements, the amplitude of background response $|<S(f_2)^{(N2)}>_{(1000)}|$ is ~ 0.17 µV (this amplitude value is represented by a black full up-triangle in Figure 6). The bandwidth-normalized standard root-mean-square deviation of background signal $\sigma(f_2)^{(N2)}$ is equal to ~ 0.082 µV/Hz$^{1/2}$ (this quantity is shown as a blue open up-triangle in Figure 4). The measured amplitude of useful signal $|<S(f_2)^{(NH3)}>_{(1000)}|$ is ~ 9.1 µV. The obtained bandwidth-normalized standard deviation for the useful signal is ~ 0.083 µV/Hz$^{1/2}$.

According to the obtained data, the fluctuations in the background signal are due to the transducer noise. The observed standard deviation of background signal $\sigma(f_2)^{(N2)}$ is approximately equal to a minimal transducer-noise level $\sigma_{f2} \approx 0.082$ µV/Hz$^{1/2}$, which can be attainable (see subsection 3.2) for a frequency $f = f_2$ in the absence of laser beam at negligibly small or moderate external disturbances. Multiple realizations of the noise ($<S(f_2)^{(1)}>_{(4)}$) and background ($<S(f_2)^{(N2)}>_{(4)}$) signals averaged over a time $\tau_{avr} = 4\tau_1$ are represented in Figure 7 as complex numbers. In the figure the black dashed line is a circle, the centre of which is located at a point near $<S(f_2)^{(1)}>_{(1000)}$. The circle radius is equal to $\sigma_{f2}$. The blue dash-dot line is a circle with the centre near the number $<S(f_2)^{(N2)}>_{(1000)}$. The radius of the circle is equal to a standard deviation of the background signal $\sigma(f_2)^{(N2)}/(4\tau_1)^{1/2} \approx \sigma(f_2)^{(N2)}$. The radiuses of both circles are approximately equal. The amplitude of background response $|<S(f_2)^{(N2)}>_{(1000)}|$ is seen from Figure 7 to be approximately equal to a quantity $2\sigma_{f2}$.

In comparison to the amplitude of useful response $|<S(f_2)^{(NH3)}>_{(1000)}|$, the quantity $|<S(f_2)^{(N2)}>_{(1000)}|$ is considerably low (by a factor of $G \approx 53.5$). The measured amplitude of the background signal is equivalent, approximately, to an absorption coefficient

$$\alpha_{bg} = \alpha^{(NH3)}C^{(NH3)}/G \approx 2.51 \cdot 10^{-5} \text{ cm}^{-1}. \quad (2)$$

### 5.1. Noise-limited minimal detectable absorption

Absorption-detection performance of the prototype cell can be specified in terms of the noise-equivalent absorption normalized to laser-beam power and detection bandwidth. This quantity corresponds to a minimal detectable absorption coefficient $\alpha_{min}$, at which the signal-to-noise ratio is equal to 1 if the laser-beam power is 1 W and the signal-averaging time is 1 s. At given values of the power $P_{on}$ and signal-to-noise ratio $SNR$ to be realized in the experiment, the coefficient $\alpha_{min}$ is found from:

$$\alpha_{min} = \alpha^{(NH3)}C^{(NH3)}P_{on} SNR^{-1}. \quad (3)$$

The signal-to-noise ratio SNR is defined as a ratio of the amplitude of useful response to the bandwidth-normalized noise level $\sigma_{f2}$. According to our measurements at $P_{on} = 6.66$ mW:

$$SNR = |<S(f_2)^{(NH3)}>_{(1000)}|/\sigma_{f2} \approx 111.0.$$

Correspondingly, the minimal detectable absorption $\alpha_{min}$ for the prototype cell is estimated to be ~ $8.07 \cdot 10^{-8}$ cm$^{-1}$ W Hz$^{-1/2}$. The minimal ammonia concentration, detectable at a typical power $P_{on}^{(typ)}$ of modulated laser beam (for the near-infrared single-mode laser diodes, this quantity is ~ 10 mW) and a signal-averaging time of 1 s, is

$$C_{min}^{(NH3)} = \alpha_{min}/(\alpha^{(NH3)}P_{on}^{(typ)}\tau_{avr}^{1/2}) \approx 23.5 \text{ ppm}. \quad (4)$$

### 5.2. Performance of gas-leak detection

The noise-limited minimal ammonia-leak rate $R_{min}^{(NH3)}$, which can be detectable in a carrier gas at a time resolution of 1 s and a typical laser-beam power $P_{on}^{(typ)}$ of 10 mW, for the cell prototype can be estimated as:

$$R_{min}^{(NH3)} = 60\pi(D/2)^2L_C C_{min}^{(NH3)} \approx 60(V/2)C_{min}^{(NH3)} \approx 3.24 \cdot 10^{-6} \text{ cm}^3/\text{min}. \quad (5)$$

Here we assume that the ammonia-containing gas carrier is blown through the cell at a flow rate (the volume of central cylindrical part of cell cavity per one second), which answers to a gas renewal inside the central cell part for the time of ~ 1 s. The ammonia content in the carrier gas to be analyzed is $C_{min}^{(NH3)}$.

The effect of background signal on the leak-detection performance can be evaluated in terms of a volumetric amount of detected gas, which enters into the cell and produces a photoacoustic response equivalent to the background signal. For ammonia blown through the central cavity part of our cell, this quantity is equal to

$$V_{bg}^{(NH3)} = \pi(D/2)^2 L_C \alpha_{bg}/\alpha^{(NH3)} \approx (V/2)\alpha_{bg}/\alpha^{(NH3)} \approx 1.7\ 10^{-7}\ cm^3. \qquad (6)$$

## 6. Discussion

### 6.1. Absorption-detection performance

Analysis of the obtained data shows that, in comparison to the macro-scale banana-shaped cell, the miniaturized prototype cell demonstrates a significantly low performance to detect absorption in gas. The noise-limited minimal detectable absorption $\alpha_{min}$ is higher for the produced cell prototype than for the cell [38] by a factor of $A \approx 404$. The high value of minimal detectable absorption for the prototype is explained, first of all, by a high level of noises to be generated by the applied MEMS-based ultrasonic transducer. We suppose also that the acoustic coupling of the transducer with the cavity of cell prototype is not optimal.

At the resonance between the modulated laser beam and acoustic $v_2$ mode, the absorption equivalent of background signal for our cell is $\alpha_{bg}/\alpha_{bg,macro} \approx 1255$ times higher than one for the cell [38]. We associate the high value of background amplitude for the produced prototype cell with a non-ideal optimization of cell design. Really, the design of the prototype cell is not properly optimized. The peak amplitude for the useful signal is realized if the laser beam is modulated at a frequency $f_2 \approx 32.9$ kHz. However, all signals, which have a negative effect on the performance of gas detection, are minimized for frequencies, which are definitely high compared to the frequency $f_2$. The cell prototype is reliably isolated from external acoustic disturbances for the range of frequencies from 35 to 50 kHz. The strong disturbances can lead to an increase in the noise-associated measurement error near the frequency $f_2$. The amplitude of parasite background signal attains its minimal value at a modulation frequency of ~ 40 kHz. This minimal value is 3.4 times lower than the background amplitude at the frequency $f_2$.

### 6.2. Performance of gas-leak detection

Despite low absorption-sensing performance, the miniaturized cell demonstrates a good capability to detect gas leaks to be emitted by a localized object. The produced cell prototype shows a significantly better performance of gas-leak detection in comparison both to the macro-scale banana-shaped cell and to the commercial portable gas-leak detectors. The noise-limited minimal detectable ammonia-leak rate $R_{min}^{(NH3)}$ is lower for the cell prototype than for the cell [38] by a factor of $V_{macro}/(V\ A) \approx 55$. The gas-leak performance of commercial hand-held halogen/hydrogen/helium sniffer leak detectors attains a rate value of ~ $10^{-5}\ cm^3/min$ [18].

The quantity $V_{bg}^{(NH3)}$, which specifies the effect of background signal on the leak-detection performance (see Eq. (6)), is smaller for the cell prototype than for the cell [38] by a factor of $V_{macro}\alpha_{bg,macro}/(V\alpha_{bg}) \approx 18.7$. Notice that, in accordance with Eq. (6), the product of the volume of central cavity part and the equivalent of background absorption is a suitable parameter for estimation of background effect for the banana-shaped photoacoustic cells irrespectively of the substance to be detected. In a sense, this parameter can be considered as a "cross section of photoacoustic background signal". For our cell prototype, this product is equal to $\alpha_{bg}(V/2) \approx 5.8\ 10^{-8}\ cm^2$.

### 6.3. Suggestions

Basing on the obtained results, we can make a few suggestions directed to enhance the performance of miniaturized resonant photoacoustic cell. First of all, we suppose that the MEMS-based ultrasonic transducer applied in the cell prototype can be successfully replaced by a sensor (for instance, by a miniature condenser microphone) of higher performance. Such a replacement can reduce significantly the noise-limited minimal detectable quantities specified by Eqs. (3) and (5).

The sensor replacement can be associated with the need to increase the cavity lengths $L_C$ and $L_{Lat}$ (these lengths are coupled by relation $L_{Lat} = L_C/2$) in order to adapt the cell for operation with a low-frequency acoustic sensor. At a constant diameter $D$, the rise of the lengths $L_C$ and $L_{Lat}$ must lead to a proportional decrease in the minimal detectable absorption $\alpha_{min}$ and concentration $C_{min}^{(NH3)}$ (here we assume that the $Q$-factor is an invariant). In accordance with Eq. (5), the noise-limited minimal gas-leak rate does not vary with the parameters $L_C$ and $L_{Lat}$.

The background effect for the resonant photoacoustic cell is an "accumulation" of imperfections of the cavity design, cell-manufacturing process and cell-alignment procedure. Certainly, the proper design optimization for a so miniaturized cell should be performed with the help of an accurate 3D numeric simulation, which takes into account features (for instance, influence of the cell sensor on the resonant acoustic standing wave) important for the consideration. The COMSOL Multiphysics software package seems to be a suitable tool, which can be applied for such purposes [31-34].

The effect of imperfections produced at a cell manufacturing is badly predictable. In general, in order to reduce such imperfections, we should provide a design simplicity and minimal human assistance at the cell production.

The alignment of the cell along the laser beam is a final factor affecting on the background signal. Certainly, the cell alignment helps to reduce the background. In our measurements, we align the cell prototype at signal-average times not longer than 1 s (a time scale comfortable to visualize a reaction of background signal to an aligning manipulation). At such times the amplitude of background signal is comparable with the noise level (see Figure 7). The reduction of the noise effect on the measurements (for instance, by a sensor replacement) will facilitate the alignment procedure and can help to reduce the background signal.

## 7. Conclusion

Thus, we have presented a resonant photoacoustic cell. The acoustic cavity for this cell is a miniaturized size-scaled imitation of the well-developed cavity shape of a macro-scale cell. We made a detailed examination of performance for a produced cell prototype of substantially reduced sizes. The advantages and drawbacks of the prototype were analyzed. Suggestions directed to enhance the performance of miniaturized photoacoustic cell are offered. In the nearest future, we will use these suggestions in order to develop a miniature cell of higher performance.

In general, we can conclude that, despite low absorption-sensing performance, the produced miniaturized cell prototype shows a high performance of gas-leak detection. A deciding factor for such an increase in leak-detection sensitivity is the reduced sizes of the prototype. The cell prototype in combination with a compact near-infrared laser can be used as a low-cost basis in order to develop a 'pocket-sized' gas-leak detector. The obtained results demonstrate a great potential for the miniaturized resonant photoacoustic cells in creating hand-held high-sensitivity laser-spectroscopy sensors of chemical compounds.

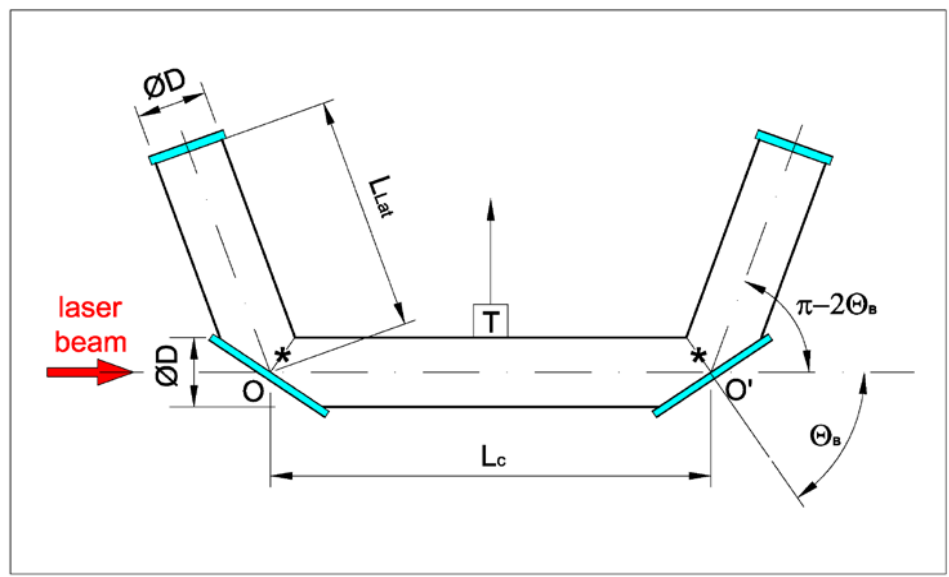

Figure 1. Design of acoustic cavity of cell: *OO'* - optical axis, $L_c$ - length of the central cylinder part, $L_{lat}$ - length of the lateral cylinder parts, *D* - cross-section diameters for the central and lateral cylinder parts, $\Theta_B$ - angle between *OO'* and the normal to the window surface, *T* - acoustic sensor. Location of the inlet/outlet gas holes is shown by stars.

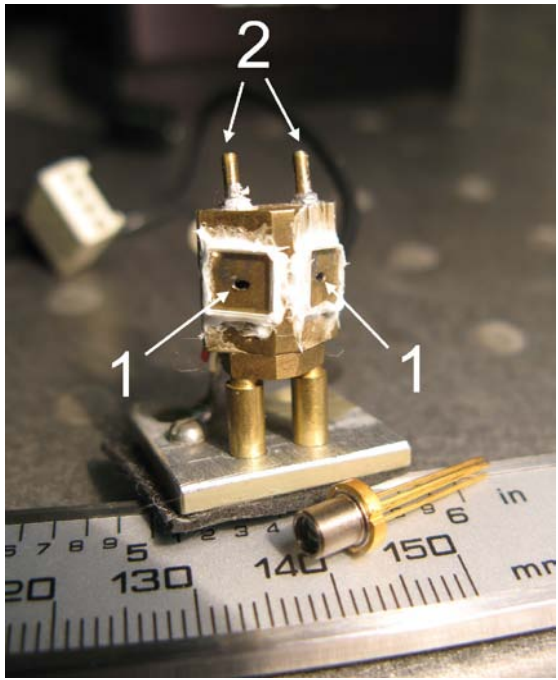

Figure 2. Photo of the produced prototype cell: 1- optical front- and back-end windows, 2 - nipples of inlet/outlet gas holes. At the photo bottom, a standard commercially available single-mode near-infrared DFB laser diode is represented

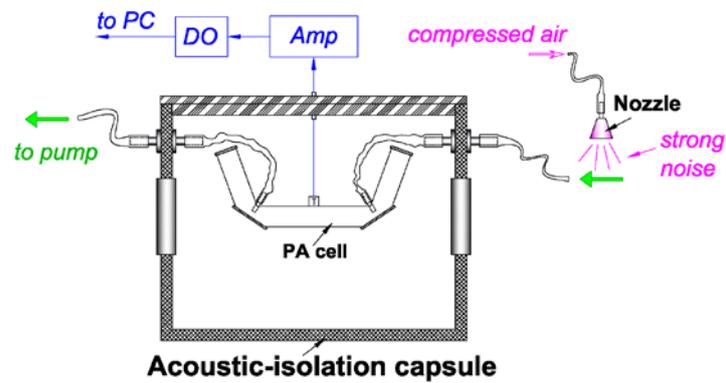

Figure 3. Experimental arrangement for estimation of the measurement error associated with transducer noise: (T) transducer; (Amp) frequency selective amplifier; (DO) digital oscilloscope.

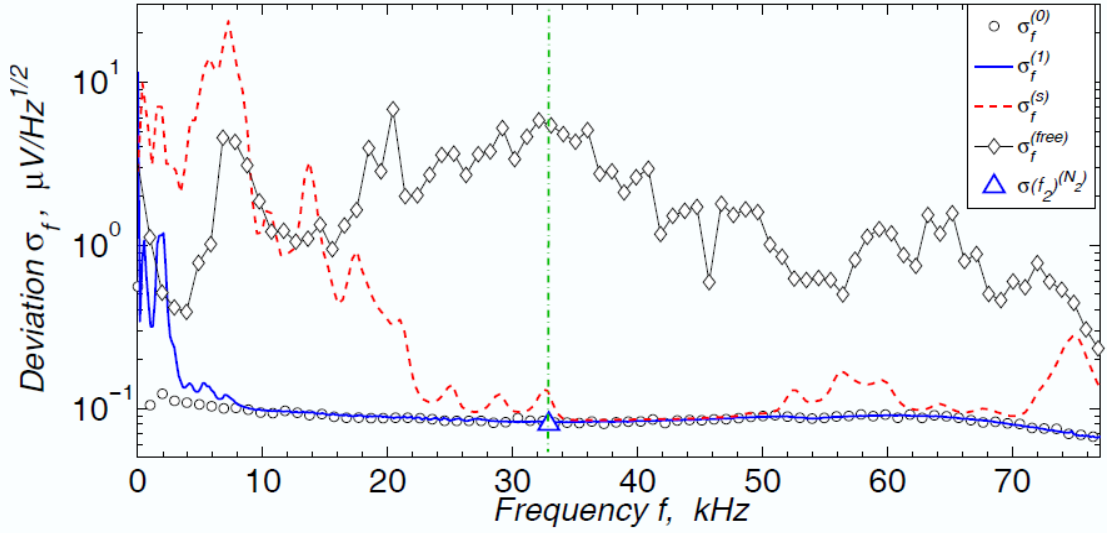

Figure 4. Frequency spectrum of bandwidth-normalized standard deviation for the Fourier transform of transducer signal detected in the absence of laser beam at negligibly small (black open circles), moderate (solid blue fat line) or strong (dashed red line) external-noise effects on the cell transducer. The solid black line with open diamonds shows a frequency spectrum of the nozzle-disturbance-initiated measurement error $\sigma_f^{(free)}$ observed for a free-spaced ultrasonic sensor SPM0204UD5 located immediately in front of the nozzle. The blue open up-triangle shows the bandwidth-normalized standard deviation $\sigma(f_2)^{(N2)}$ obtained for the laser-initiated background signal at a standard day-to-day noise level in the laboratory room (see subsection 4.2). The vertical green dash-dot line answers to location of the resonance frequency $f_2 \approx$ 32.9 kHz on the abscissa axis.

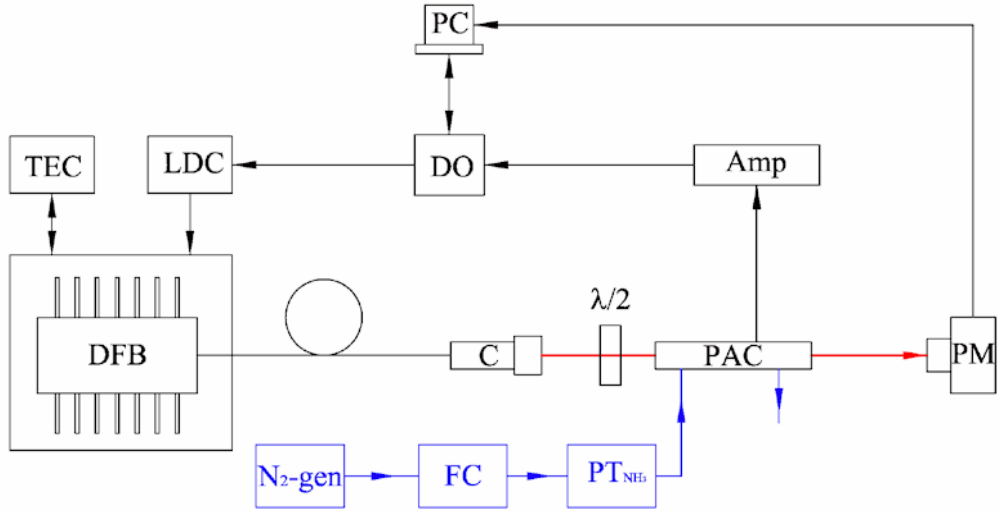

Figure 5. Experimental setup for analysis of laser-initiated photoacoustic signals: (N$_2$-gen) nitrogen generator; (FC) gas-flow controller; (PT$_{NH3}$) box calibrated permeation tube with ammonia; (LDC) current controller; (TED) temperature controller; (DFB) fiber-pigtailed laser diode; (C) fiber collimator; ($\lambda$/2) half-wave plate; (PAC) photoacoustic cell; (PM) power meter; (Amp) frequency selective amplifier; (DO) digital oscilloscope; (PC) personal computer.

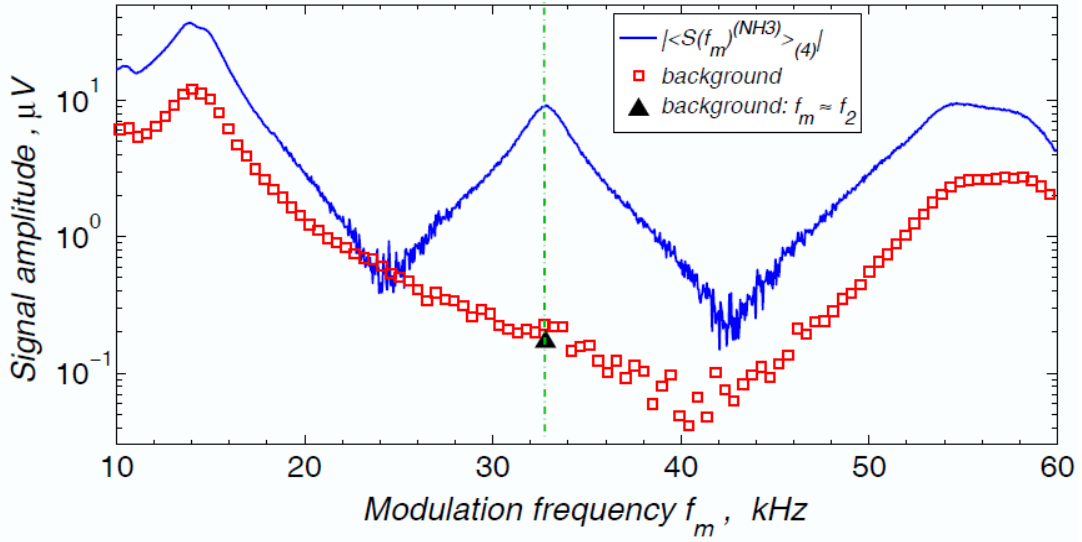

Figure 6. Amplitude-frequency dependence of laser-initiated photoacoustic responses detected inside the cell prototype at two distinct gas flows. The red open squares show the dependence of background-signal amplitude $|<S(f_m)^{(N2)}>_{(40)}|$ observed for a flow of pure nitrogen at $\tau_{avr} = 40\tau_1 \approx 10.5$ s. The blue solid line demonstrates the dependence of signal amplitude $|<S(f_m)^{(NH3)}>_{(4)}|$ obtained at a nitrogen flow admixed by ammonia ($\tau_{avr} = 4\tau_1 \approx 1.05$ s). The black full up-triangle gives the amplitude of background response ($\tau_{avr} = 1000\tau_1 \approx 262$ s) at $f_m = f_2$ and $P_{on} \approx 6.66$ mW. The vertical green dash-dot line answers to location of resonance frequency $f_2$ on the abscissa axis.

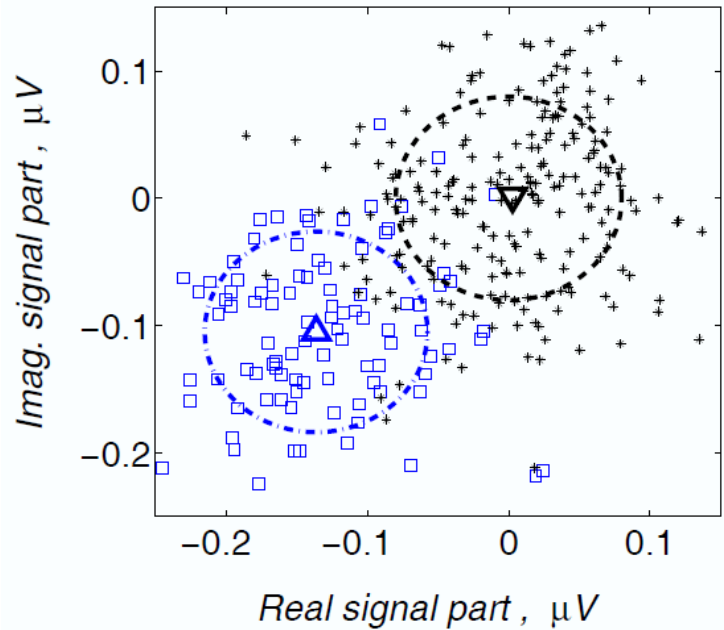

Figure 7. Realizations of the noise (black stars) and background (open blue squares) signals represented as complex-valued numbers at a signal-averaging time $\tau_{avr} = 4\tau_1 \approx 1.05$ s. A down-triangle and up-triangle give location on the complex plane for, correspondingly, the noise and background signals, which are averaged over a time scale $\tau_{avr} = 1000\tau_1 \approx 262$ s.